\documentclass[manuscript,nonacm]{acmart}

\AtBeginDocument{%
  }

\begin{document}

\title{Same Performance, Hidden Bias: Evaluating Hypothesis- and Recommendation-Driven AI}

\author{Michaela Benk}
\affiliation{%
  \institution{University of Zurich}
  \city{Zurich}
  \country{Switzerland}}
\email{michaela.benk@uzh.ch}

\author{Tim Miller}
\affiliation{%
  \institution{The University of Queensland}
  \city{Brisbane}
  \country{Australia}}
\email{timothy.miller@uq.edu.au}

\renewcommand{\shortauthors}{Benk}

\begin{abstract}
The HCI community commonly evaluates decision support systems based on whether they improve task performance or promote appropriate user reliance. In this work, we look beyond decision outcomes to examine the process through which users develop decision-making strategies. Through a web-based experiment (N = 290) comparing recommendation-driven and hypothesis-driven interaction designs, and using Signal Detection Theory as a theoretical framework, we show that even when performance remains identical, recommendation-driven designs lower participants’ thresholds for sufficient evidence and introduce a ``hidden bias'' in their judgments, resulting in a shifted distribution of errors. Furthermore, we find that experts are just as susceptible to these systemic shifts as novices. We conclude by advocating for a shift in focus: prioritizing decision processes and the preservation of stable evidence standards over performance and reliance alone.

\end{abstract}




\maketitle

\section{Introduction}\label{introduction}

Consider two forensic analysts, Laura and Sergio, both using AI-based decision support systems to identify suspects in a large case involving thousands of digital files. Laura uses a recommendation-driven system: it suggests a match, highlights evidence, and provides a high confidence score.
Laura reviews the system's output and agrees with the recommendation; she trusts the output, as the highlighted segments seem convincing. Meanwhile, Sergio uses a hypothesis-driven system: it provides no recommendation, but offers tools to explore ambiguities and alternative interpretations of the data. After careful deliberation, he also identifies the same suspect.

On the surface, the outcomes are identical. Both analysts reach the same conclusion. Both trust their respective system's outputs. And both perform accurately on this case. However, the decision support paradigms of the two systems each foster qualitatively different decision-making strategies. The `hidden bias' lies in how each system affects their cognitive process used to assess evidence, and the resulting error patterns.

As noted by \citet{Miller2023ExplainableAI}, decision-makers in AI-assisted settings engage in two distinct but interrelated processes: (a) evaluating options and outcomes (``Is this a match?'') and (b) assessing the soundness of the DSS advice (``Should I trust this output?''). In hypothesis-driven settings, the analyst's error patterns will primarily reflect their personal threshold used to evaluate evidence. In recommendation-driven settings, however, the evaluation of the recommendation conflates with the evaluation of the task: the AI's ``match'' affects the threshold for what the analyst considers sufficient evidence and subsequently, the errors they make. 

While HCI evaluations frequently focus on performance (e.g., accuracy) and reliance (e.g., agreement with the AI's output)\cite{Bucinca-et-al_2021_Forcing, langer2024effective, Chen2023UnderstandingTR, schemmer2022should, nourani2021anchoring, Vasconcelos2022ExplanationsCR}, they fail to reveal the hidden bias: whether and how the interaction design shifts the evidence threshold. This shift has profound implications for what \citet{langer2024effective} term 'effective human oversight.' They argue that humans must employ decision strategies that reliably detect errors to act as effective safeguards, a task that depends on how the users sets their threshold for accepting evidence before making a decision. By analyzing how design alters these thresholds, we can better understand whether an interface supports or undermines the human's ability to balance errors.

\paragraph{\textbf{Signal Detection Theory as a Lens.}}
Signal Detection Theory (SDT) provides a formal framework to decouple these processes by separating sensitivity ($d'$), the ability to distinguish signal from noise, from the criterion ($c$), the evidence threshold that reflects the decision strategy \cite{stanislaw1999sdt}. In high-stakes domains, a ``liberal'' decision strategy (negative $c$ values/lower threshold) increases the risk of false accusations, even if accuracy remains high. In our context, a liberal shift in $c$ indicates that the analyst requires less evidence to confirm a suspect. Conversely, a ``conservative'' shift (positive values) might delay critical interventions. SDT has only recently been highlighted as an effective method to model human oversight in AI-assisted decision making \cite{langer2024effective, Elder_2024}. 

\paragraph{\textbf{Contribution}.} Prior work has recognized the need to look at specific error patterns and highlighted how oversight effectiveness diminishes when users fail to distinguish AI errors from correct outputs \cite{langer2024effective}. Our work extends this by demonstrating how specific interaction paradigms, namely recommendation-driven vs. hypothesis-driven designs, systematically alter evidence thresholds. Using SDT as a robust theoretical framework, we demonstrate that:

\begin{itemize}
    \item \textbf{Hidden bias:} While participants across conditions show similar overall accuracy, recommendation-driven designs shift users toward a lower threshold of evidence required. Conversely, hypothesis-driven approaches maintain similar thresholds as having no AI assistance.
    \item \textbf{Adaptation:} We observe a divergence in how users adapt their strategies. Participants across conditions improve their performance over time, but only those in the recommendation-driven condition systematically adapt their evidence thresholds. 
    \item \textbf{Expertise:} Our findings reveal that domain expertise does not provide immunity against the `hidden bias' that the interaction design can induce. 
\end{itemize}

We hope to foster a discussion in the research community about moving beyond traditional performance and reliance metrics and toward designing and evaluating AI systems in terms of the hidden bias they may elicit. 

\section{Method}

\subsection{Task and Materials}
To study decision strategies in a controlled, yet domain-general setting, we designed a decision-making task inspired by the forensic investigation vignette from Section \ref{introduction}. The task preserves three key properties: (a) comparable difficulty for humans and AI, (b) binary decisions enabling decision-theoretic analysis, and (c) known ground truth.

We adapted the \textit{OnlyConnectWall} dataset \cite{alavi2024large}, which consists of word association puzzles containing misleading cues (“red herrings”). Participants were shown sets of 16 words and asked to judge whether a specific theme was present or not. Errors could occur as false positives (accepting an impostor theme) or false negatives (rejecting a true theme), allowing analysis of SDT metrics. 

AI assistance was simulated using pre-generated outputs from GPT-4, calibrated to achieve 70\% accuracy, reflecting realistic but imperfect decision support behavior observed in prior work \cite{Bansal2021DoesTW, Bucinca-et-al_2021_Forcing}. All stimuli and system outputs were fixed across conditions.

\subsection{Study Design}
\subsubsection{Experimental Conditions}

Our overarching hypothesis posits that the interaction design influences people's decision strategy. To test this hypothesis, participants were randomly assigned to one of five conditions:
\begin{itemize}
    \item \textbf{Control}: no AI assistance
    \item \textbf{Recommendation-driven (RD)}: AI provides a binary recommendation.
    \item \textbf{Recommendation + explanation (RDX)}: recommendation plus explanatory information and uncertainty cues.
    \item \textbf{Recommendation + hypothesis exploration (RDH)}: recommendation with interactive exploration of alternative interpretations.
    \item \textbf{Hypothesis-driven (HD)}: participants explored evidence and uncertainty without a recommendation.
\end{itemize}

The conditions varied only in interaction design; task content was identical.

\subsubsection{Participants and Procedure}
We recruited 320 UK-based participants from Prolific; after exclusions, 290 remained (mean age = 42, 49\% female). Participants completed a training phase with feedback followed by a testing phase without feedback. In each trial, they judged whether a theme was present and identified supporting words if applicable. 

\subsubsection{Measures and Analysis}
Our main dependent variables were sensitivity ($d'$) and decision criterion ($c$). There are diverse possible adaptations of SDT. In this work, sensitivity is calculated as $d' = Z(H) - Z(F)$, where $H$ is the hit rate and $F$ is the false alarm rate. The criterion is calculated as $c = -0.5[Z(H) + Z(F)]$. We consider a user's ability to detect a theme to be a ``hit'' and failure to do so a ``miss''.

Accuracy, decision time, and self-reported cognitive load were included as secondary measures. We used mixed-effects models to account for repeated measures across phases and participants. Statistical details are provided in the Appendix.

\section{Results}\label{results}

\subsection{Task Performance}
\textbf{Efficiency.} Condition significantly impacted decision time ($p < .001$). Participants in HD spent the longest (M=48.5s), while RD was the fastest (M=28.9s).

\textbf{Accuracy.} Brier scores revealed no significant differences in overall accuracy between conditions ($p > .05$), suggesting that while interaction design altered the process, it did not significantly change the final performance outcomes.

\textbf{Decision Criterion ($c$).} Model 1 ($R^2 = .391$) revealed a significant, conservative baseline intercept ($\hat{\beta} = 1.16, p < .001$). Compared to the Control, we observed significantly more \textit{liberal} (lower) decision thresholds for RD ($\hat{\beta} = -0.19$, SE = 0.06, $p = .002$), DX ($\hat{\beta} = -0.34$, SE = 0.06, $p < .001$), and RDH ($\hat{\beta} = -0.21$, SE = 0.06, $p < .001$). Crucially, the HD condition did not significantly differ from the control group ($\hat{\beta} = -0.09$, SE = 0.06, $p = .137$). 
Furthermore, post-hoc tests showed that participants in RD ($p = .0015$) and RDX ($p < .0001$) significantly lowered their thresholds between the training and testing phases. No such temporal threshold adaptation was observed in the HD or Control groups.

\textbf{Sensitivity ($d'$).} Model 2 ($R^2 = .253$) showed no significant main effect of Condition on $d'$. However, all conditions showed significantly higher sensitivity in the test phase compared to training ($p < .001$), with the largest improvement occurring in the HD condition.

\section{Discussion}\label{discussion}

Our findings demonstrate that the choice of interaction design actively shapes the decision strategies users employ, even when performance remains constant.

{\textbf{Interaction Design effects.}} Our results show that in the absence of recommendations, participants maintain a conservative threshold ($c$), mirroring the independent standard expected of Sergio in our vignette. However, recommendation-driven designs systematically push users toward a liberal bias. As \citet{langer2024effective} argue, \textit{effective human oversight} relies on the user's ability to maintain a stable criterion to detect AI inaccuracies or unfairness \cite{langer2024effective}. 

{\textbf{Adaptation over time.}} We found that only RD and RDX users adapted their thresholds over time. This suggests that recommendations act as a ``moving anchor'' that can reshape the user's strategy as they observe the system. The lack of sensitivity ($d'$) improvement in RD compared to HD suggests that decision makers tend to adapt their strategy selection to the type of decision aids available in such a way as to reduce effort \cite{todd1991decision}, where the presence of a recommendation discourages the cognitive engagement necessary to refine one's independent discrimination ability.

{\textbf{No expertise invariance.}} Notably, while task knowledge and age significantly predicted sensitivity, they had no effect on the decision criterion. This suggests that expert users are just as susceptible to the `hidden bias' as novices; expertise improves discrimination but does not immunize the user against the systemic threshold shifts induced by the interface.

{\textbf{Design implications.}} Focusing on performance or reliance measures fails to capture the `hidden bias' of AI interfaces. For high-stakes oversight, we advocate for designs that prioritize the maintenance of stable, independent evidence thresholds. SDT provides a robust evaluative framework to ensure that AI assistance supports, rather than undermines, the human's role as a reliable safeguard in high-risk socio-technical systems.

{\textbf{Future avenues}.} We are pursuing several extensions to this work. First, while our 20-trial design enabled controlled measurement of decision patterns (median completion: 35 minutes), replication with larger samples and additional contexts will establish the robustness and generalizability of our findings. Second, our static task environment prioritized internal validity, but real-world AI use is becoming increasingly interactive. Investigating how reliance strategies evolve when users iteratively engage with explanations represents a key next step for ecological validity.
Finally, we focused on a single explanation method with lay participants. Comparative studies across explanation types (e.g., feature importance, counterfactuals, examples) and expert populations would reveal whether our results reflect general principles or are context-dependent.

\section{Conclusion}\label{conclusion}
This research investigates how different AI design paradigms influence human decision-making strategies, revealing a `hidden bias' that is not visible in conventional metrics. We propose that the HCI research community move beyond evaluating AI through the lens of performance outcomes and instead adopt formal frameworks like SDT to understand the strategic interplay between interface paradigms and human cognition.
\bibliographystyle{ACM-Reference-Format}


\newpage

\appendix

\section{Study Design}

\begin{figure}[h!]
    \centering
    \includegraphics[width=1\linewidth]{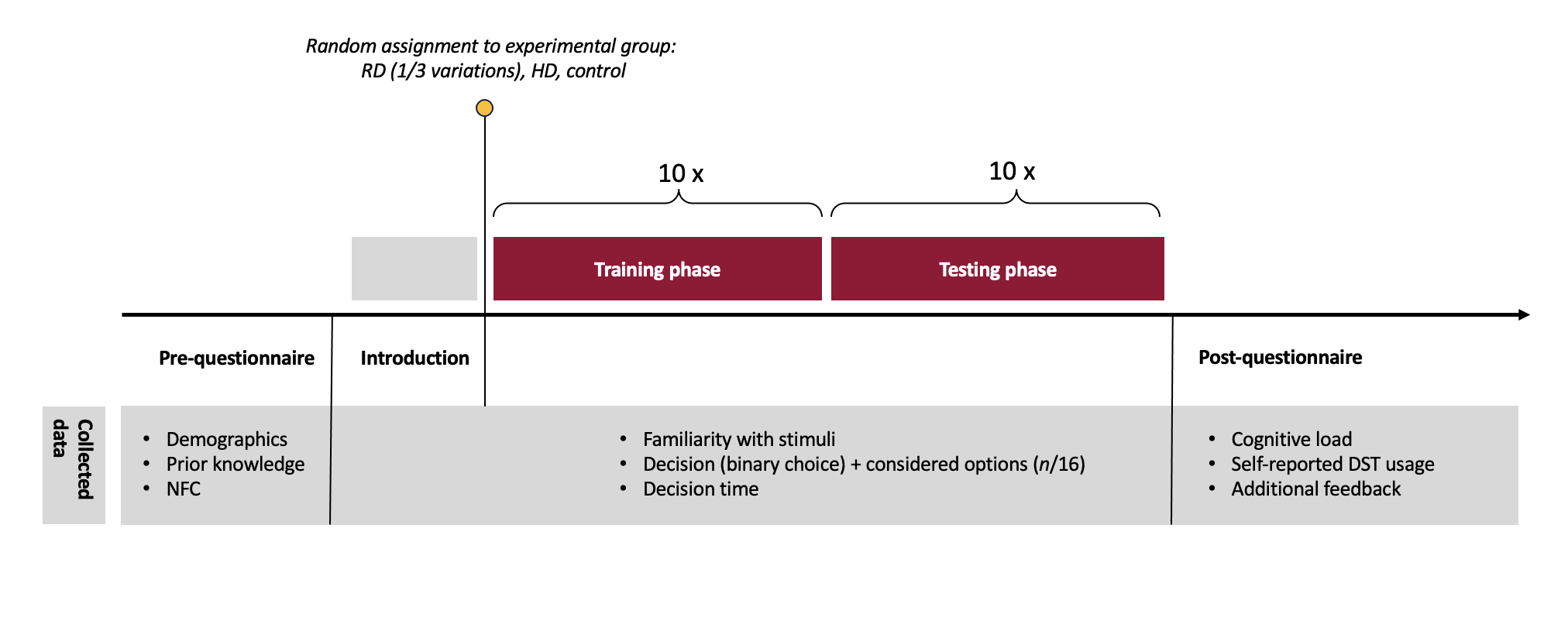}
    \caption{Experimental Protocol Overview. The study pipeline comprises a pre-experimental assessment of individual differences (Prior Knowledge, NFC), followed by random assignment into experimental conditions. The core task is split into 10 training and 10 testing trials where behavioral metrics are captured. The session concludes with a post-hoc evaluation of cognitive load and tool engagement.}
    \label{fig:procedure}
\end{figure}

\section{Stimuli}

\begin{figure}[h!]
    \centering
    \includegraphics[width=1\linewidth]{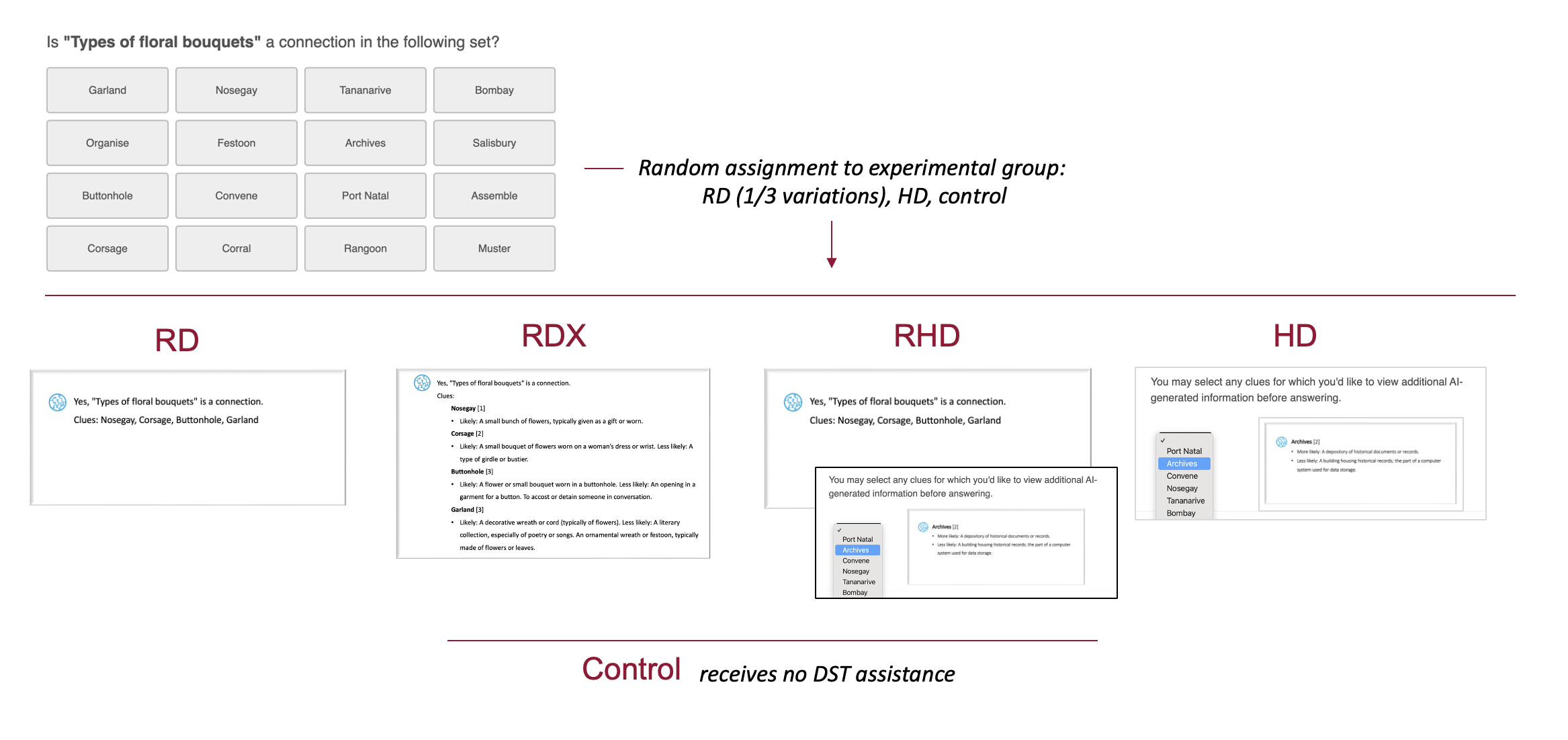}
    \caption{Example of task stimuli, using the OnlyConnect dataset, designed to evaluate participants' ability to determine the presence of a theme.}
    \label{fig:stimuli}
\end{figure}

\newpage

\section{Analysis}

\begin{figure}[h]
    \centering
    \includegraphics[width=\linewidth]{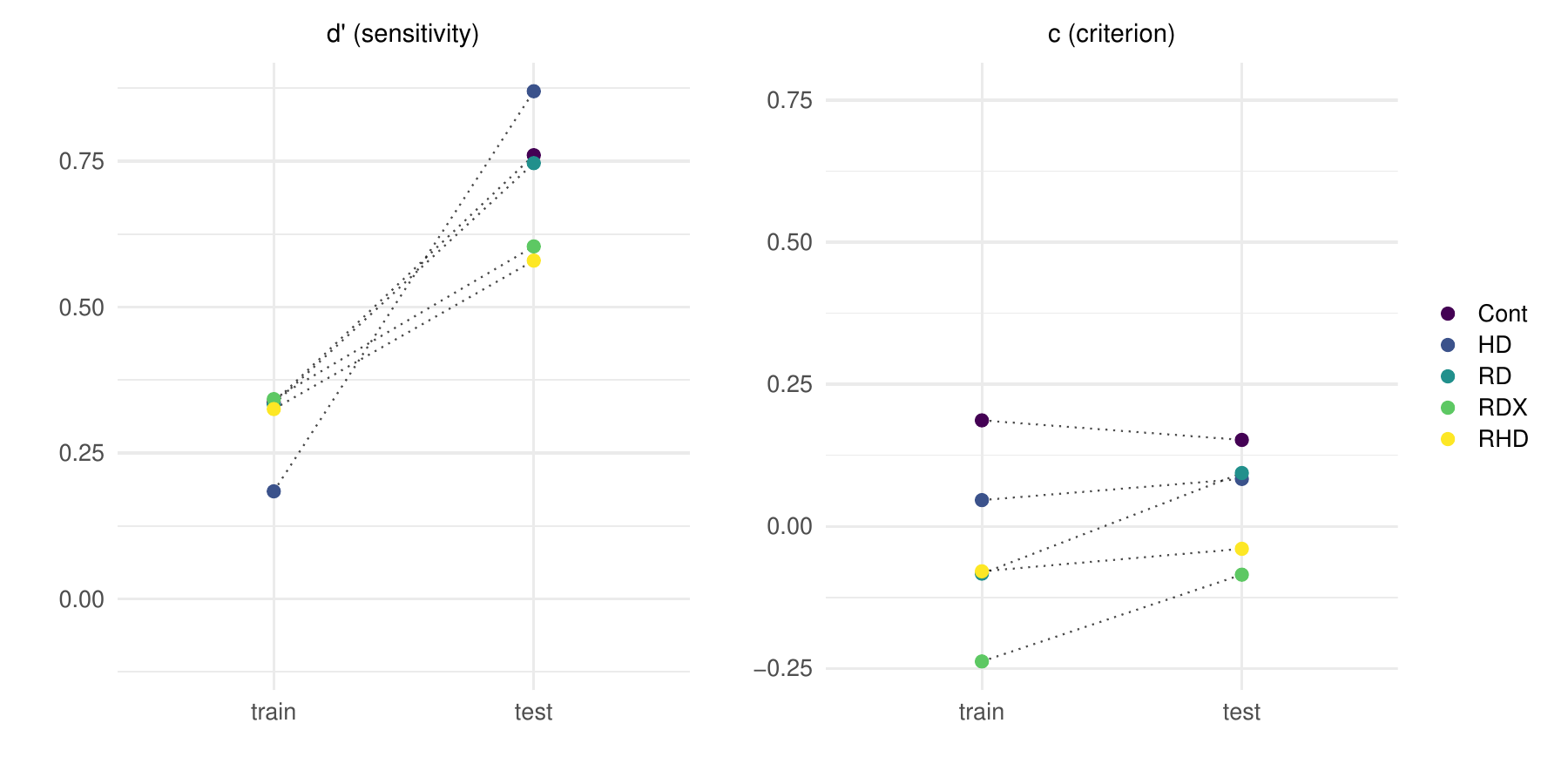}
    \caption{(left) Sensitivity ($d'$) and (right) decision criterion ($c$) across two study phases. While sensitivity improved for all participants, criterion shifts were specific to recommendation-driven designs.}
    \label{fig:sdt}
\end{figure}


\newpage

\end{document}